# Platform Competition with User-Generated Content

Bohan ZHANG[a,1]
[a]*School of Sciences, Chang'an University, Xi'an, China*
ORCiD ID: Bohan Zhang https://orcid.org/0009-0007-8218-4390

**Abstract.** This paper develops a theoretical model of platform competition where user-generated content (UGC) quality arises endogenously from the composition of the user base. Users differ in their relative preferences for content quality and network size, and platforms compete by choosing advertising intensity, which affects user utility through perceived quality. We characterize equilibrium platform choice, identifying conditions under which equilibria are stable. The model captures how platforms' strategic decisions shape user allocation and market outcomes, including coexistence and dominance scenarios. We consider two types of equilibria in advertising levels: Nash equilibria and Stackelberg equilibria, and discuss the industry and policy implications of our results.

## 1. Introduction

User-generated content (UGC), defined as material that a platform sources from its own end-users, forms the backbone of most social media platforms. Platforms such as YouTube, Facebook, Instagram, and TikTok owe much of their success to the vast amounts of content generated by their users [1].

Traditionally, economic models of platform competition have placed significant emphasis on content quantity, which is often proxied by user base size or content volume [2]. This focus aligns with the notion that a larger volume of content attracts more users. However, this perspective overlooks a critical dimension of platform competitiveness, which is content quality. High-quality content not only captivates audiences but also fosters trust and credibility, which are essential for sustained user retention. For instance, platforms like Xiaohongshu (RedNote) have demonstrated that prioritizing user feedback and co-creation can lead to more authentic and engaging content, enhancing user satisfaction and driving sales for brands.[2] Similarly, YouTube's shift from human-curated content to algorithmically-generated highlights, driven by popularity metrics, has sparked discussions about the quality of content being promoted.[3]

---

[1] Corresponding Author: Bohan Zhang, zhangbohan@chd.edu.cn.
[2] https://www.voguebusiness.com/consumers/how-xiaohongshu-is-supercharging-the-co-creation-economy-in-china
[3] https://www.wired.com/story/social-media-human-curation

This also creates an opportunity for smaller platforms to compete by offering higher-quality content. However, unlike traditional media platforms where quality is controlled by the platform [3], UGC platforms face the challenge of content quality being endogenously determined by user contributions. This user-driven content creation model means that the quality of content is heavily influenced by the platform's user base. Platforms' strategic choices, particularly regarding advertising intensity, play a pivotal role in shaping content quality. Aggressive advertising strategies can lead to a decrease in perceived quality, which can then lead to the loss of high-quality users.

This paper develops a theoretical model where content quality arises endogenously from user contributions and is shaped by platforms' advertising intensity. Users differ in how they value content quality relative to network size, and their collective choices determine both the distribution of quality and market shares. The contribution of the paper is threefold. First, we extend the literature on platform competition by explicitly modeling the feedback loop between advertising, user base composition, and content quality. Second, we characterize the equilibrium properties and introduce a refinement process to select among multiple, plausible outcomes. Third, we show that platform competition with UGC results in the dominant platform occupying the whole market. These insights provide not only theoretical advances but also practical guidance for platform operators and policymakers concerned with digital market concentration.

## 2. Literature Review

### *2.1. Platforms and Network Effects*

The economic model of two-sided platforms emphasizes the role of network effects in shaping competitive outcomes. Early studies such as [4] and [5] show how user adoption depends on installed base and cross-side externalities. Belleflamme and Peitz [2] synthesize previous work in the literature, demonstrating that positive feedback often results in tipping and winner-takes-all dynamics. These models typically focus on network effects generated from user base size and either ignore user quality or implicitly assume that more users always lead to better content. Akerlof's "lemons" model [6] captures how quality and participation interact under uncertainty, and concludes that due to adverse selection no equilibrium can exist, however, in Akerlof's model, the quality of a good provided by a seller is treated as fixed, and cannot be changed strategically.

### *2.2. Quality Competition in Media and Digital Markets*

There are studies in the literature that treats quality as a strategic variable in market competition. For instance, Battaggion and Drufuca [3] analyze quality competition and market entry within two-sided platforms, demonstrating that the threat of a new entrant can disrupt the market equilibrium by compelling incumbents to reduce their level of quality differentiation. In a similar vein, Gabszewicz et al. [7] model the broadcasting industry as a sequential game, where platforms first choose their programming profiles and subsequently set their advertising levels. Their analysis reveals that as viewers' aversion to advertising intensifies, platforms are pushed toward more homogenized content, making targeted "niche" strategies less viable. While these works adeptly illustrate the critical trade-off between monetization and user experience, they share a common underlying assumption: that quality is a variable directly and unilaterally controlled by the platform, rather than being an emergent property of user activity and production.

*2.3. User-Generated Content Models*

UGC platforms introduce a fundamentally different mechanism: quality emerges from the user base itself, and often cannot be directly controlled by the platform. Luca [1] investigates factors that influence the quality of UGC and points out that platform competition with UGC "can lead to multiple equilibria, in which some platforms are filled with excellent content and other equally good platforms are virtual ghost towns". Zhang and Sarvary [8] develop a Hotelling-style model with a circular city similar to Salop [9] where UGC differentiation arises through user heterogeneity. Deng et al. [10] show empirically that editorial reviews interact with UGC to influence perceived value. Although these studies recognize the importance of UGC, they generally treat its distribution as exogenous and do not link it explicitly to platform advertising strategies.

*2.4. Research Gap and Contribution*

In summary, prior work has separately examined network effects, quality competition, and UGC. Yet little attention has been paid to how platform decisions shape UGC quality endogenously through user heterogeneity and advertising intensity. Our paper fills this gap by integrating these elements into a unified model. In doing so, we explain how competition can collapse in simultaneous play, why sequential commitment generates first-mover dominance, and what trade-offs platforms face between monetization and content quality. This contributes to both theory and practice by offering new insights into digital platform competition, market concentration, and managerial strategy.

## 3. Model Setup

*3.1. User Utility*

We consider a consumer population of measure 1 and two competing platforms, indexed by $j = 1, 2$. Users contribute content within their chosen platforms. The content quality of each user $i$ is characterized by a parameter $q_i > 0$. Both platforms are assumed to face the same operating costs, which are normalized to 0.

A user's utility from joining a platform depends on two components: the quality of content available on the platform and the size of its user base, reflecting network effects. At the same time, advertising, while profitable for the platform, diminishes user utility by lowering the perceived quality of the content. Specifically, we assume that the utility of user $i$ derives from platform $j$ is given by

$$U_{ij} = \alpha_i^q \left(Q_j - \gamma a_j\right) + \alpha_i^n n_j, \tag{1}$$

where

$$Q_j = \mathbb{E}^*[q_i \mid \text{User } i \text{ is on platform } j] \tag{2}$$

denotes the average content quality on platform $j$, $a_j$ is the advertising intensity, and $\gamma > 0$ captures the extent to which advertising reduces perceived quality (assumed the same across all users). Note here we define the conditional expectation operator in an unconventional way: if the set which is conditioned on has measure 0, then we set $\mathbb{E}^*[q_i \mid \text{User } i \text{ is on platform } j] = -\infty$; otherwise it is defined as usual. The asterisk is to differentiate it from the usual conditional expectation operator, which we denote by $E[\cdot \mid \cdot]$. The interpretation of this is that when a platform has no users, the average quality of that platform is not defined; any other platform with users will be considered of higher quality, and therefore, preferred to it. This may seem like an "abuse of notation," but it greatly simplifies our results. The parameter $n_j$ denotes the measure of users on platform $j$, while $\alpha_i^q$ and $\alpha_i^n$ represent user $i$'s marginal utility from content quality and network size, respectively.

Since utility is ordinal, the utility function can be normalized without loss of generality. First, we set the unit of advertising intensity such that $\gamma = 1$. Then we divide through by $\alpha_i^q$ and define $\beta_i = \alpha_i^n / \alpha_i^q$ (which can be interpreted as user $i$'s valuation of network size relative to content quality). This gives us

$$U_{ij} = Q_j + \beta_i n_j - a_j, \tag{3}$$

which follows closely the specification in [8] and allows for a tractable analysis. We assume that $\beta_i \sim U[0, 1]$.

*3.2. Platform Choice*

Users choose the platform that maximizes their utility. Let $X = [0, 1]$ be the set of all possible values of $\beta_i$, and $\mu$ be the Lebesgue measure on $[0, 1]$. Throughout this section we assume that $a_1$ and $a_2$ are fixed. First we characterize an allocation.

**Definition 1** (Allocation). An *allocation* is a pair of sets $(A, B)$ such that

$$A \cup B = X, \quad A \cap B = \varnothing, \tag{4}$$

that is, every agent chooses either platform 1 ($A$) or platform 2 ($B$). Note that we do not require $A$ and $B$ be non-empty. Also we have the identity $\mu(A) + \mu(B) = 1$.

To characterize an equilibrium allocation in our model, we utilize a generalized version of *fulfilled-expectations equilibrium* [8, 11]. In its classic formulation, a fulfilled-expectations equilibrium involves a mapping from consumers' expected market shares to realized market shares, under the assumption that consumer utility depends only on firms' market shares [8]. The equilibrium is characterized as a fixed point of this mapping—i.e., expectations are confirmed in equilibrium. We generalize this concept to our model, where consumer utility takes a different form.

**Definition 2** (Realized Allocation). Given an *expected allocation* $(A^e, B^e)$, a *realized allocation* $(A^r, B^r)$ is an allocation such that for all $\beta_a \in A^r$,

$$\mathbb{E}^*[q_i \mid \beta_i \in A^e] + \beta_a \mu(A^e) - a_1 \geq \mathbb{E}^*[q_i \mid \beta_i \in B^e] + \beta_a \mu(B^e) - a_2, \tag{5}$$

and for all $\beta_b \in B^r$,

$$\mathbb{E}^*[q_i \mid \beta_i \in A^e] + \beta_b \mu(A^e) - a_1 \leq \mathbb{E}^*[q_i \mid \beta_i \in B^e] + \beta_b \mu(B^e) - a_2. \tag{6}$$

These inequalities ensure that an allocation is consistent with utility maximization under the consumers' expectations. Note that the realized allocation may not be unique.

Let the average content quality of a user with valuation $\beta$ be

$$q(\beta) = \mathbb{E}[q_i \mid \beta_i = \beta]. \tag{7}$$

A natural assumption is that $q(\beta)$ is monotonically decreasing, so that higher-$\beta$ users (low relative valuation of quality) tend to provide lower-quality posts. In our model, we set $q(\beta) = q_m - \lambda\beta$, where $\lambda > 0$ represents the strength of the relationship between user's valuation of quality and quality of content contribution.

**Definition 3** (Equilibrium Allocation). An *equilibrium allocation* is a pair $(A, B)$ such that, when $(A, B)$ is the expected allocation, it is also the realized allocation (i.e., the expectations are self-fulfilling).

Intuitively, users with a high relative valuation of network effects (i.e., higher $\beta_i$) tend to prefer the platform with a larger user base, and users with a lower valuation of network effects tend to prefer the platform offering higher quality. This suggests the existence of a threshold type such that all users with $\beta_i$ above the threshold choose one platform, while those below it choose the other. We formalize and prove this intuition in the following proposition.

**Proposition 1**. If $(A^r, B^r)$ is a realized allocation where $\mu(A^r) \neq \mu(B^r)$, then one of the following must hold:

$$\forall \beta_a \in A^r,\ \beta_b \in B^r,\ \beta_a > \beta_b, \quad \text{or} \quad \forall \beta_a \in A^r,\ \beta_b \in B^r,\ \beta_a < \beta_b. \tag{8}$$

*Proof.* Let $\beta_a \in A^r$, $\beta_b \in B^r$. From the equilibrium conditions, we have:

$$\begin{aligned} (\mathbb{E}^*[q_i \mid \beta_i \in A^e] + \beta_a\mu(A^e) - a_1) - (\mathbb{E}^*[q_i \mid \beta_i \in B^e] + \beta_a\mu(B^e) - a_2) &\geq 0, \\ (\mathbb{E}^*[q_i \mid \beta_i \in A^e] + \beta_b\mu(A^e) - a_1) - (\mathbb{E}^*[q_i \mid \beta_i \in B^e] + \beta_b\mu(B^e) - a_2) &\leq 0. \end{aligned} \tag{9}$$

Therefore,

$$(\beta_a - \beta_b)\big(\mu(A^e) - \mu(B^e)\big) \geq 0. \tag{10}$$

This implies that the sign of $\beta_a - \beta_b$ agrees with the sign of $\mu(A^e) - \mu(B^e)$, and since $A \cap B = \varnothing$, $\beta_a - \beta_b \neq 0$. So if $\mu(A^e) > \mu(B^e)$, we must have $\beta_a > \beta_b$, and if $\mu(A^e) < \mu(B^e)$, then $\beta_a < \beta_b$. It follows that either all $\beta_a > \beta_b$ or all $\beta_a < \beta_b$, which is to be proven. □

We exclude the case where $\mu(A) = \mu(B)$ for reasons to be explained later; to summarize, an equilibrium under the condition $\mu(A) = \mu(B)$ is not stable.

**Corollary 1**. By the completeness property of real numbers, every realized allocation (and therefore, equilibrium allocation) produces a unique cutoff $\tilde{\beta} \in [0, 1/2]$ such that

$$\tilde{\beta} = \begin{cases} \sup A = \inf B & \text{if } \mu(A) > \mu(B), \\ \inf A = \sup B & \text{if } \mu(A) < \mu(B). \end{cases} \tag{11}$$

*Note 1*. We will refer to both the equilibrium allocation $(A, B)$ and the corresponding cutoff $\tilde{\beta}$ as the *equilibrium*. This dual usage will not cause confusion because one is a pair of sets and the other a real number. Also from now on we do not distinguish between two allocations $(A_1, B_1)$ and $(A_2, B_2)$ with the same market shares, i.e., $\mu(A_1) = \mu(A_2)$, $\mu(B_1) = \mu(B_2)$. It is obvious that given the market share of platform 1 in a realized allocation $\mu(A^r) = n_1^r$, then the allocation $(A^r, B^r)$ is identified up to a set with measure 0.

**Definition 4** (Equilibrium Market Shares). Let $(A, B)$ be an equilibrium allocation. Define

$$n_1^* = \mu(A), \quad n_2^* = \mu(B). \tag{12}$$

We call $(n_1^*, n_2^*)$ the *equilibrium market shares*. Since $\mu(A) + \mu(B) = 1$, it suffices to specify $n_1^*$ only.

Given the equilibrium market share of platform 1 (denoted as $n_1^*$), the relationship between $n_1^*$ and the equilibrium cutoff, $\tilde{\beta}^*$, can be written as

$$n_1^* = \begin{cases} 1 - \tilde{\beta}^*, & \text{if } n_1^* > \dfrac{1}{2}, \\ \tilde{\beta}^*, & \text{if } n_1^* < \dfrac{1}{2}. \end{cases} \tag{13}$$

**Proposition 2** (Uniqueness of Realized Allocation). Given an expected allocation $(A^e, B^e)$, a realized allocation is unique in the sense that for any two realized allocations $(A_1^r, B_1^r)$ and $(A_2^r, B_2^r)$,

$$\mu(A_1^r) = \mu(A_2^r), \quad \mu(B_1^r) = \mu(B_2^r). \tag{14}$$

*Proof.* Suppose, to the contrary, that there exist two realized allocations from the same expected allocation $(A^e, B^e)$, namely $(A_1^r, B_1^r)$ and $(A_2^r, B_2^r)$, with $\mu(A_1^r) \neq \mu(A_2^r)$. Without loss of generality, assume that $\mu(A_1^r) > \mu(A_2^r)$. Then

$$A_1^r - A_2^r = A_1^r \cap B_2^r \neq \varnothing. \tag{15}$$

Let $\beta$ be an element of $A_1^r - A_2^r$. Then $\beta$ must simultaneously satisfy (5) and (6), which means that

$$(\mathbb{E}^*[q_i \mid \beta_i \in A^e] + \beta\mu(A^e) - a_1) - (\mathbb{E}^*[q_i \mid \beta_i \in B^e] + \beta\mu(B^e) - a_2) = 0. \tag{16}$$

Since this is a linear equation with respect to $\beta$, where the coefficient of $\beta$ is not 0, the solution set contains at most one element (if the solution falls within $[0, 1]$ then there is one element, otherwise there is none), thus $\mu(A_1^r - A_2^r) = 0$. Similarly $\mu(A_2^r - A_1^r) = 0$.[4] Note that

$$|\mu(A_1^r) - \mu(A_2^r)| \leq \mu(A_1^r - A_2^r) + \mu(A_2^r - A_1^r) = 0, \tag{17}$$

which proves that $\mu(A_1^r) = \mu(A_2^r)$; therefore $\mu(B_1^r) = \mu(B_2^r)$ as well. □

With the result of Proposition 2, we can define the mapping from the expected market shares to the realized market shares. Specifically, the realized share of platform 1, $n_1^r$, is unique when the expected share of platform 1, $n_1^e$, is given:

$$n_1^r = \Gamma(n_1^e). \tag{18}$$

Define

$$\Gamma(n_1^e) = \begin{cases} 0, & \text{if } n_1^e = 0, \\ \min\left\{1, \dfrac{[Q_2^e(n_1^e) - Q_1^e(n_1^e)] - (a_2 - a_1)}{2n_1^e - 1}\right\}, & \text{if } 0 < n_1^e < \dfrac{1}{2}, \\ \max\left\{0, 1 - \dfrac{[Q_2^e(n_1^e) - Q_1^e(n_1^e)] - (a_2 - a_1)}{2n_1^e - 1}\right\}, & \text{if } \dfrac{1}{2} < n_1^e < 1, \\ 1, & \text{if } n_1^e = 1. \end{cases} \tag{19}$$

where the expected average content quality on the platforms are

---

[4] It might be the case that

$$A_2^r - A_1^r = A_2^r \cap B_1^r = \varnothing.$$

Under such cases, (16) is vacuously true. Though one can also isolate this special case to ensure that $A_2^r - A_1^r \neq \varnothing$.

$$
\begin{aligned}
Q_1^e &= \begin{cases} \dfrac{\int_{\tilde{\beta}}^{1} q\,(\beta)\,\mathrm{d}\beta}{1-\tilde{\beta}}, & \text{if } n_1^e > \dfrac{1}{2}, \\[2ex] \dfrac{\int_{0}^{\tilde{\beta}} q\,(\beta)\,\mathrm{d}\beta}{\tilde{\beta}}, & \text{if } n_1^e < \dfrac{1}{2}, \end{cases} \\
Q_2^e &= \begin{cases} \dfrac{\int_{0}^{\tilde{\beta}} q\,(\beta)\,\mathrm{d}\beta}{\tilde{\beta}}, & \text{if } n_1^e > \dfrac{1}{2}, \\[2ex] \dfrac{\int_{\tilde{\beta}}^{1} q\,(\beta)\,\mathrm{d}\beta}{1-\tilde{\beta}}, & \text{if } n_1^e < \dfrac{1}{2}, \end{cases}
\end{aligned} \tag{20}
$$

and

$$
n_1^e = \begin{cases} 1-\tilde{\beta}, & \text{if } n_1^e > \dfrac{1}{2}, \\[1ex] \tilde{\beta}, & \text{if } n_1^e < \dfrac{1}{2}. \end{cases} \tag{21}
$$

Plugging in $q(\beta) = q_m - \lambda\beta$, we can simplify (20) to:

$$
\Gamma(n_1^e) = \begin{cases} 0, & \text{if } n_1^e = 0, \\[1ex] \min\left\{1, \dfrac{-\dfrac{\lambda}{2} - (a_2 - a_1)}{2n_1^e - 1}\right\}, & \text{if } 0 < n_1^e < \dfrac{1}{2}, \\[3ex] \max\left\{0, 1 - \dfrac{\dfrac{\lambda}{2} - (a_2 - a_1)}{2n_1^e - 1}\right\}, & \text{if } \dfrac{1}{2} < n_1^e < 1, \\[1ex] 1, & \text{if } n_1^e = 1. \end{cases} \tag{22}
$$

The interior equilibrium value of $\tilde{\beta}$ must satisfy one of the following conditions:

$$
\left(1 - 2\tilde{\beta}\right)\tilde{\beta} = a_1 - a_2 + \frac{\lambda}{2}, \quad (n_1 > n_2) \tag{23}
$$

or

$$
\left(2\tilde{\beta} - 1\right)\tilde{\beta} = a_1 - a_2 - \frac{\lambda}{2}. \quad (n_1 < n_2) \tag{24}
$$

where $\tilde{\beta} \in (0,1/2)$. Thus, the condition for the existence of an interior equilibrium in which either platform 1 or platform 2 dominates is given by:

$$
-\frac{\lambda}{2} < a_1 - a_2 \le \frac{1}{8} - \frac{\lambda}{2}, \quad (n_1 > n_2) \tag{25}
$$

or

$$
-\frac{1}{8} + \frac{\lambda}{2} \le a_1 - a_2 < \frac{\lambda}{2}. \quad (n_1 < n_2). \tag{26}
$$

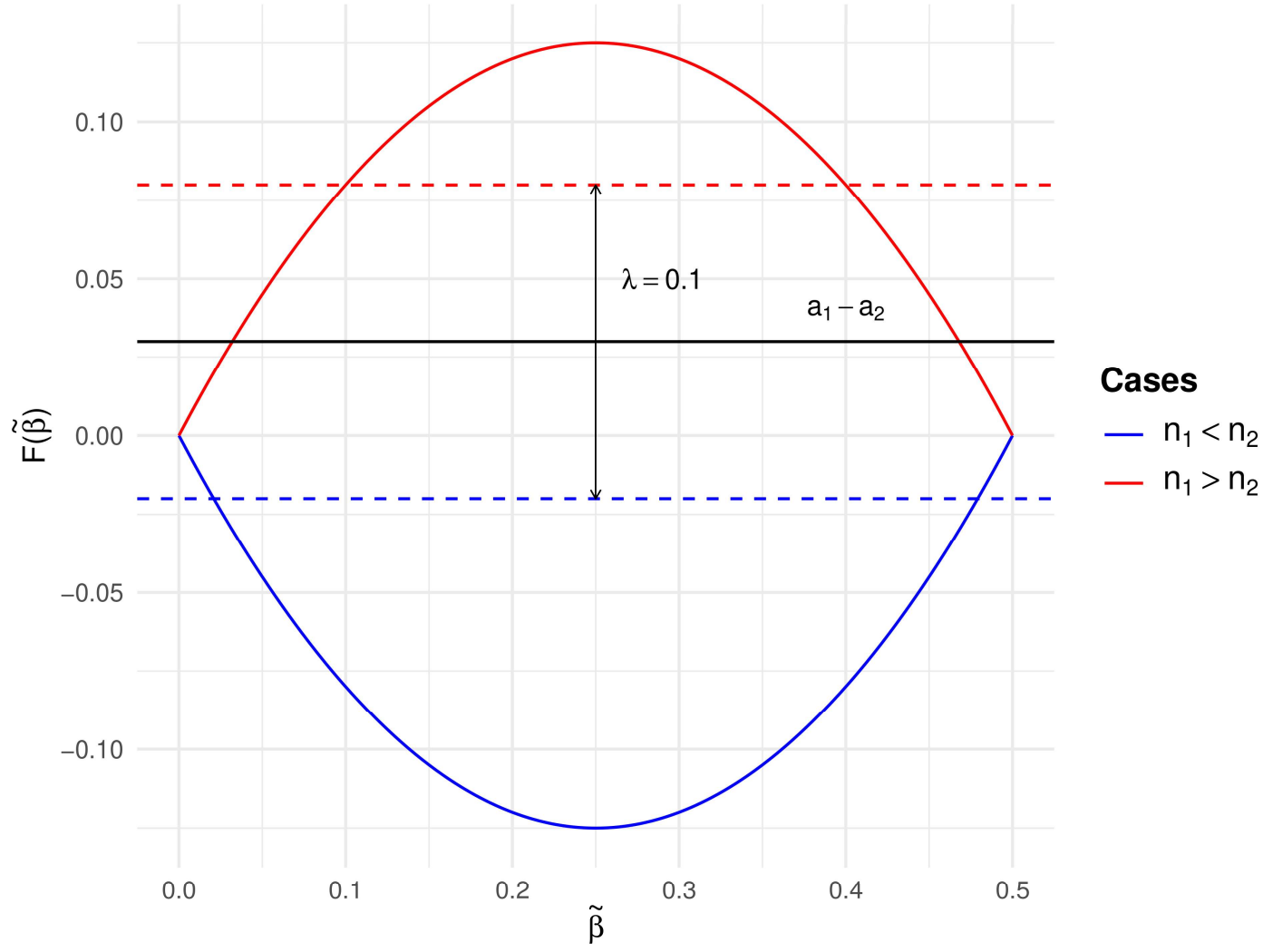

**Figure 1.** $\lambda = 0.10$

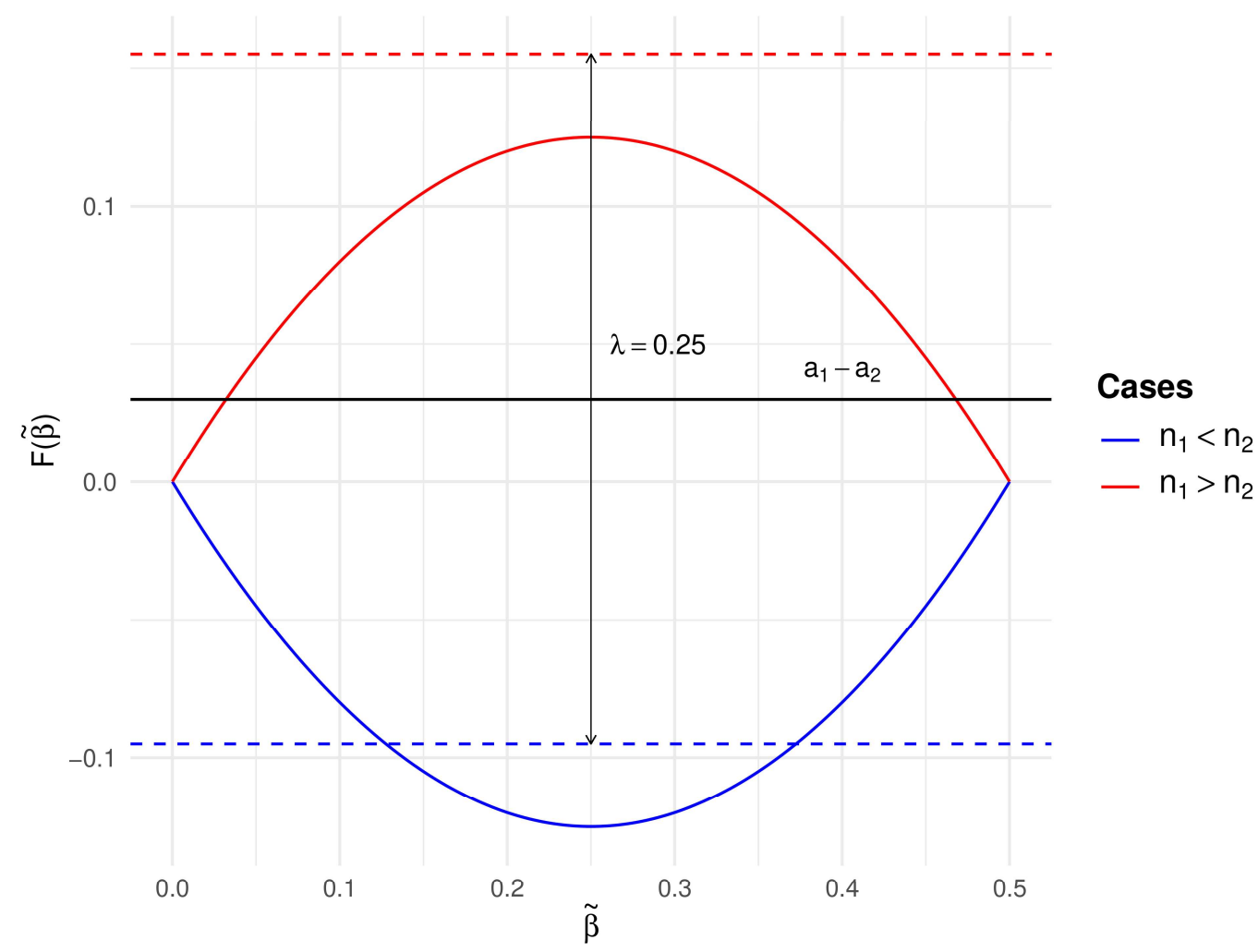

**Figure 2.** $\lambda = 0.25$

These conditions can be easily verified with graphs: In Figure 1 and 2 the red and blue parabola represent the functions $F(\tilde{\beta}) = (1 - 2\tilde{\beta})\tilde{\beta}$ and $F(\tilde{\beta}) = (2\tilde{\beta} - 1)\tilde{\beta}$, defined on $(0,1/2)$. The black horizontal line is $y = a_1 - a_2$, and the red and blue dashed lines are $y = a_1 - a_2 + \lambda/2$ and $y = a_1 - a_2 - \lambda/2$. Note that the gap between the two dashed lines is $\lambda$. For an equilibrium where $n_1$ dominates to exist, the red dashed line must intersect the red parabola. Therefore $a_1 - a_2 + \lambda/2$ must be between 0 and $1/8$ (the maximum height of the parabola). Similarly, for an equilibrium where $n_2$ dominates to exist, $a_1 - a_2 - \lambda/2$ must be between $-1/8$ and 0.

**Corollary 2**. Equilibria where $n_1$ dominates and equilibria where $n_2$ dominates can coexist if and only if

$$-\frac{1}{8}+\frac{\lambda}{2}<\frac{1}{8}-\frac{\lambda}{2}, \tag{27}$$

that is,

$$\lambda<\frac{1}{4}. \tag{28}$$

However, the case where $\lambda \geq 1/4$ poses a problem: when $a_1 - a_2$ is close to 0, no interior equilibrium can exist. This is contrary to many real-world scenarios, where two platforms exhibit significant quality differences driven solely by their user bases. Therefore for our analysis we assume $\lambda < 1/4$.

*3.3. Equilibrium Refinement*

In many models involving rational expectation, it is typical for multiple equilibria to coexist [12], which is also the case in our model (note that (24) and (25) are quadratic with respect to $\tilde{\beta}$). However, refinement criteria can be applied to select a subset of equilibria that are more "natural" than others. Specifically, we use the concept of stability of exclude certain equilibria.

**Definition 5** (Stable Equilibrium). Let $(A, B)$ be an equilibrium allocation, and $(A^e, B^e)$ be an expected allocation which satisfies the condition (8). We say that an equilibrium is stable, if there exists an $\varepsilon > 0$, such that

$$|n_1^* - \mu(A^e)| < \varepsilon \Longrightarrow \lim_{k\to\infty} n_1^k = n_1^* \tag{29}$$

where $n_1^k$ is the sequence given by

$$n_1^{k+1} = \Gamma(n_1^k), \quad n_1^0 = \mu(A^e). \tag{30}$$

Note that we assert condition (8) to be true for the expected allocation, so that we can identify the allocation up to a set with measure 0 using only the market share of platform 1, like in the case of realized allocations. This is not a restrictive assertion, since (8) is always true for realized allocations; the condition will automatically hold after one iteration, and we can redefine this realized allocation from the first iteration as the expected allocation.

**Proposition 3**. If the condition (23) or (24) yields multiple (unequal) equilibrium, then only the smaller one is stable.

*Proof.* We only prove the result for (23), and the proof for (24) is similar. Assume $\varepsilon$ is sufficiently small (so that it falls within the same branch). Rewrite the iteration in terms of $\tilde{\beta}$:

$$\tilde{\beta}_{n+1} = \frac{a_1 - a_2 + \frac{\lambda}{2}}{1 - 2\tilde{\beta}_n}. \tag{31}$$

Let $c = a_1 - a_2 + \frac{\lambda}{2}$, then $c \in (0, 1/8)$. An equilibrium $\beta^*$ is (locally) stable if $|f'(\beta^*)| < 1$, where $f(x) = c/(1-2x)$.

Denote the two fixed points by

$$\beta_1^* = \frac{1-\sqrt{1-8c}}{4}, \quad \beta_2^* = \frac{1+\sqrt{1-8c}}{4}. \tag{32}$$

The derivative at $\beta_1^*$ is

$$f'(\beta_1^*) = \frac{8c}{\left(1+\sqrt{1-8c}\right)^2}. \tag{33}$$

Since $0 < 8c < 1$ and $\left(1+\sqrt{1-8c}\right)^2 > 1$ for $c \in (0,1/8)$, $|f'(\beta_1^*)| < 1$. Thus $\beta_1^*$ is locally stable.

The derivative at $\beta_2^*$ is

$$f'(\beta_2^*) = \frac{8c}{\left(1-\sqrt{1-8c}\right)^2}. \tag{34}$$

Let $u = \sqrt{1-8c}$. Then

$$f'(\beta_2^*) = \frac{8c}{(1-u)^2} = \frac{1-u^2}{(1-u)^2}, \tag{35}$$

Since $u \in (0,1)$, $2u^2 - 2u < 0$. Therefore

$$\frac{1-u^2}{(1-u)^2} > 1. \tag{36}$$

Thus

$$|f'(\beta_2^*)| > 1, \tag{37}$$

which means that $\beta_2^*$ is unstable. □

**Proposition 4**. If $a_1 - a_2 > \lambda/2$, then the boundary equilibrium where the platform 2 dominates is stable. If $a_1 - a_2 < -\lambda/2$, then the boundary equilibrium where the platform 1 dominates is stable.

*Proof.* We only prove the case where $a_1 - a_2 > \lambda/2$, since the other case is similar (or can be obtained by symmetry). Let $n_1^e = \varepsilon$. If

$$\frac{\lambda}{2} + \beta(2\varepsilon - 1) - (a_1 - a_2) < 0 \tag{38}$$

for all $\beta \in [0,1]$, then the equilibrium is stable. For sufficiently small $\varepsilon$, $2\varepsilon - 1$ is negative, therefore (38) holds. Thus the boundary equilibrium where platform 2 dominates is stable. □

Aside from the criterion of stability, we also introduce the concept of a focal platform (as in [13]). We assume that consumers play the equilibrium in which they join the focal platform when multiple stable equilibria exist.[5] Without loss of generality, we assume that platform 1 is focal.

---

[5] This is also justified by the iteration as defined in (30). For example, suppose platform 1 monopolizes the market prior to the entry of platform 2. If there exists a stable interior equilibrium in which platform 1 dominates, and we initialize the iteration with $n_1^e = 1 - \varepsilon$ for sufficiently small $\varepsilon$, the dynamic process converges to that equilibrium. This justifies selecting the dominant platform as the focal point in such settings.

*3.4. Profit Maximization*

Platforms generate revenue by selling advertising space. Following [8], we assume that a platform's profit is proportional to both its advertising level and the size of its user base. Ignoring the proportionality constant, we can write

$$\pi_j = a_j n_j. \tag{39}$$

Platforms choose their advertising intensity $a_j$ to maximize their profit, subject to the equilibrium conditions governing user choice and quality.

Summarizing the results from Proposition 3 and 4, we specify the profit functions as follows:

$$\begin{aligned}
\pi_1(a_1, a_2) &= \begin{cases} \frac{a_1}{4}\left(3 + \sqrt{1 - 8\left(a_1 - a_2 + \frac{\lambda}{2}\right)}\right), & -\frac{\lambda}{2} \le a_1 - a_2 \le \frac{1}{8} - \frac{\lambda}{2}, \\ a_1, & a_1 - a_2 < -\frac{\lambda}{2}, \\ 0, & a_1 - a_2 > \frac{1}{8} - \frac{\lambda}{2}. \end{cases} \\
\pi_2(a_1, a_2) &= \begin{cases} \frac{a_2}{4}\left(1 - \sqrt{1 - 8\left(a_1 - a_2 + \frac{\lambda}{2}\right)}\right), & -\frac{\lambda}{2} \le a_1 - a_2 \le \frac{1}{8} - \frac{\lambda}{2}, \\ 0, & a_1 - a_2 < -\frac{\lambda}{2}, \\ a_2, & a_1 - a_2 > \frac{1}{8} - \frac{\lambda}{2}. \end{cases}
\end{aligned} \tag{40}$$

## 4. Characterization of Equilibria

*4.1. Nash Equilibria*

We first solve for Nash equilibria, where two platforms compete simultaneously. We try to find an interior solution first. Consider the first-order conditions:

$$\frac{\partial \pi_1}{\partial a_1} = 0, \quad \frac{\partial \pi_2}{\partial a_2} = 0. \tag{41}$$

Instead of directly solving the FOC, we define an auxiliary variable

$$u = 1 - 8\left(a_1 - a_2 + \frac{\lambda}{2}\right). \tag{42}$$

The partial derivatives of $u$ with respect to $a_1$ and $a_2$ are given by

$$\frac{\partial u}{\partial a_1} = -8, \quad \frac{\partial u}{\partial a_2} = 8. \tag{43}$$

Next, introduce the auxiliary functions

$$\hat{\pi}_1(a_1, a_2, u) = \frac{a}{4}\left(3 + \sqrt{u}\right), \quad \hat{\pi}_2(a_1, a_2, u) = \frac{a}{4}\left(1 - \sqrt{u}\right). \tag{44}$$

The derivatives of the original profit functions can be rewritten as

$$\frac{\partial \pi_1}{\partial a_1} = \frac{\partial \hat{\pi}_1}{\partial a_1} + \frac{\partial \hat{\pi}_1}{\partial u}\frac{\partial u}{\partial a_1} = \frac{3+\sqrt{u}}{4} - \frac{a_1}{\sqrt{u}} = 0, \tag{45}$$

$$\frac{\partial \pi_2}{\partial a_2} = \frac{\partial \hat{\pi}_2}{\partial a_2} + \frac{\partial \hat{\pi}_2}{\partial u}\frac{\partial u}{\partial a_2} = \frac{1-\sqrt{u}}{4} - \frac{a_2}{\sqrt{u}} = 0. \tag{46}$$

Let $c = \sqrt{u}$. Then

$$\frac{\partial \pi_1}{\partial a_1} = \frac{3+c}{4} - \frac{a_1}{c} = 0, \tag{47}$$

$$\frac{\partial \pi_2}{\partial a_2} = \frac{1-c}{4} - \frac{a_2}{c} = 0. \tag{48}$$

Solving for $a_1$ and $a_2$, we obtain:

$$a_1 = \frac{3}{4}c + \frac{1}{4}c^2, \quad a_2 = \frac{1}{4}c - \frac{1}{4}c^2. \tag{49}$$

Note that

$$a_1 + a_2 = c, \quad a_1 - a_2 = \frac{1}{2}c + \frac{1}{2}c^2. \tag{50}$$

From the definition of $u$, we also have:

$$a_1 - a_2 = \frac{1-c^2}{8} - \frac{\lambda}{2}. \tag{51}$$

Equating the two expressions for $a_1 - a_2$ gives

$$c^* = \frac{-2+\sqrt{9-20\lambda}}{5}. \tag{52}$$

We only take the positive solution because $\sqrt{u} > 0$. Plug $c^*$ back into (49) gives

$$a_1^* = \frac{11\sqrt{9-20\lambda} - 20\lambda - 17}{100}, \quad a_2^* = \frac{9\sqrt{9-20\lambda} + 20\lambda - 23}{100}. \tag{53}$$

However, we can show that this is not a Nash equilibrium. Fix $a_2 = a_2^*$ and consider platform 1's profit:

$$\pi_1(a_1, a_2^*) = \begin{cases} \frac{a_1}{4}\left(3 + \sqrt{1 - 8\left(a_1 - a_2^* + \frac{\lambda}{2}\right)}\right), & -\frac{\lambda}{2} \le a_1 - a_2^* \le \frac{1}{8} - \frac{\lambda}{2}, \\ a_1, & a_1 - a_2^* < -\frac{\lambda}{2}, \\ 0, & a_1 - a_2^* > \frac{1}{8} - \frac{\lambda}{2}. \end{cases} \tag{54}$$

First, we prove that $a_1^*$ is optimal within the first branch. Fix $a_2^*$ and $\lambda$, and let

$$f(a_1) = \pi_1(a_1, a_2^*; \lambda) \tag{55}$$

and let $u(a_1) = 1 - 8(a_1 - a_2^* + \lambda/2)$, we have

$$\begin{aligned} f'(a_1) &= \frac{3+\sqrt{u}}{4} - \frac{a_1}{\sqrt{u}}, \\ f''(a_1) &= -\frac{2}{\sqrt{u}} - \frac{4a_1}{u^{\frac{3}{2}}} < 0. \end{aligned} \tag{56}$$

therefore $a_1^*$ must be a unique maximum point within the first branch.

Next, we prove that the second branch is strictly dominated. Note that the maximum profit obtainable within the second branch occurs at the boundary, where $a_1 - a_2^* = -\frac{\lambda}{2}$. Since this profit is equal to that of the first branch at the same value of $a_1$, it must be dominated by the optimal profit in the first branch. Thus $a_1^*$ is the global maximizer of platform 1's profit.
The same process can be repeated for $a_2$. Fix $a_1 = a_1^*$, platform 2's profit is

$$\pi_2(a_1^*, a_2) = \begin{cases} \frac{a_2}{4}\left(1 - \sqrt{1 - 8\left(a_1^* - a_2 + \frac{\lambda}{2}\right)}\right), & -\frac{\lambda}{2} \le a_1^* - a_2 \le \frac{1}{8} - \frac{\lambda}{2}, \\ 0, & a_1^* - a_2 < -\frac{\lambda}{2}, \\ a_2, & a_1^* - a_2 > \frac{1}{8} - \frac{\lambda}{2}. \end{cases} \quad (57)$$

Define $u(a_2) = 1 - 8(a_1^* - a_2 + \lambda/2)$. Let

$$f_2(a_2) = \pi_2(a_1^*, a_2, \lambda). \quad (58)$$

Then for $\lambda < 1/4$,

$$f_2''(a_2^*) = -\frac{2}{\sqrt{u}} + \frac{4a_2^*}{u^{\frac{3}{2}}} > 0, \quad (59)$$

therefore $a_2^*$ cannot be a maximum point within the first branch. However, in the third branch, maximum profit is $-\frac{1}{8} + \frac{\lambda}{2} + a_1^*$, which is greater than that of the first branch. Therefore $(a_1^*, a_2^*)$ cannot be a Nash equilibrium.

But from the analysis above, it is easy to see that $(1/8 - \lambda/2, 0)$ must be a Nash equilibrium, since the maximum profit platform 2 can attain is zero when the platform 1 sets its advertising level to $1/8 - \lambda/2$ (no matter in which branch platform 2 plays).

### *4.2. Stackelberg Equilibria*

Now we turn to consider a different kind of equilibria, namely the Stackelberg equilibria. To be consistent with previous assumptions, assume that platform 1 (the focal platform) moves first, and platform 2 follows. First we notice that for platform 2, for any value of $a_1$, the profit in the first branch is always dominated by the optimal profit in the third branch, that is, $a_1 - \frac{1}{8} + \frac{\lambda}{2}$. Therefore platform 2 will always play in the third branch, where platform 1 gets zero profit. Therefore the optimal strategy of platform 1 is to set $a_1^* = \frac{1}{8} - \frac{\lambda}{2}$, effectively making the optimal profit of platform 2 zero. Therefore there exists a Stackelberg equilibrium

$$\left(\frac{1}{8} - \frac{\lambda}{2}, 0\right). \quad (60)$$

## 5. Discussion

### *5.1. Economic Intuition and Interpretation of Results*

The core result of our model is the existence of a Nash equilibrium at the strategy profile $(a_1^*, a_2^*) = (1/8 - \lambda/2, 0)$. In this equilibrium, the dominant platform retains a sustainable advertising level, while the other platform is effectively blockaded from the market, earning zero profit.

The Nash equilibrium can also be seen as the result of a dynamic adjustment process. Suppose the two platforms start at any strategy profile $(a_1, a_2)$. Platform 2 will try to take over the market by undercutting its advertising level such that $a_1 - a_2 > 1/8 - \lambda/2$. However, platform 1 will also undercut its advertising level by setting $a_1$ such that $a_1 - a_2 \leq 1/8 - \lambda/2$. The process goes on until $a_2$ reaches 0, and $a_1$ reaches $1/8 - \lambda/2$.

Notably, this equilibrium coincides exactly with the Stackelberg outcome where platform 1 is the leader, which indicates that the advantage is so potent that the dominant platform can occupy the entire market even without a sequential-move commitment. The mere potential for platform 1 to play this dominant strategy in a simultaneous game is enough to deter entry entirely.

We have assumed that there is a positive correlation between a user's relative valuation of quality and their quality of contributed content. This assumption finds strong empirical support in real-world platforms. In hobbyist forums, academic platforms, or creative communities, "hardcore" users typically possess deeper knowledge, stronger curation standards, and produce more valuable content than casual users. These quality-sensitive users also exhibit higher valuation for quality content.

It is also worth noticing that for both the interior solution and the boundary solution, an increase in $\lambda$ would result in a decrease in the optimal advertising level of platform 1, $a_1^*$ (and a decrease in $a_2^*$ for the interior solution). This relationship reveals a fundamental strategic constraint: when user-generated content quality is highly responsive to user composition (high $\lambda$), the platform's optimal strategy shifts toward lighter monetization to preserve its most valuable asset: the high-quality user segment that drives the platform's content ecosystem. The dominant platform must temper its monetization strategy to avoid driving away the high-quality contributors who are essential to its competitive advantage.

### *5.2. Illustrative Example: EJMR vs. EconSpark/EconTrack*

Though it is difficult to find a direct real-world example due to the rareness of extreme market tipping, the logic of our model can be extended to non-commercial platforms to explain the persistent dominance of forums with significant negative externalities. A pertinent example is the failure of professional alternatives like the American Economic Association's (AEA) EconSpark/EconTrack to displace the established but often-toxic academic forum Economics Job Market Rumors (EJMR, now XJMR). In this context, we can reinterpret the "advertising intensity" of our model to represent anything that causes disutility to the user (in this case, toxic content).

Though EJMR is not strategically choosing the level of toxic content, our model can still explain why alternatives like EconSpark/EconTrack, which attempted to compete by offering a platform with lower level of toxic content (also high-quality, professional environment), eventually failed. As our model shows, a new entrant offering higher quality cannot succeed if the dominant platform's advertising intensity is sufficiently low. The "advertising intensity" (toxicity) on EJMR, while significant, is not high enough to allow coexistence of another platform, hence the eventual failure of EconSpark/EconTrack.

### *5.3. Comparison with Literature*

Our model builds on and extends several strands of the platform competition literature. First, relative to classic analyses of network effects [2, 4, 5], our framework introduces a new channel: endogenous quality shaped by user heterogeneity. Whereas prior studies emphasize how larger user bases increase value through sheer scale, we show that the composition of the user base is equally important, since the departure of quality-sensitive users can trigger a negative feedback loop that erodes a platform's attractiveness despite its scale.

Second, our results differ from models of quality competition in media markets [3, 7], where quality is assumed to be directly controlled by firms. In our setting, platforms cannot unilaterally set quality but influence it indirectly through advertising intensity. This mechanism generates new strategic trade-offs: platforms must weigh monetization against the retention of high-quality contributors who sustain their competitive position. Third, compared to prior models of UGC platforms [8, 10], we explicitly link platform strategies to the endogenous distribution of content quality. Earlier work recognizes that UGC quality matters but treats its distribution as exogenous. By making quality an outcome of user composition, our model explicit models the feedback loop between user composition and content quality which is an intrinsic part of platform competition with UGC.

### *5.4. Managerial Implications*

Our results highlight important takeaways for platform managers. First, aggressive monetization through advertising can backfire when user-generated content quality is highly sensitive to user composition. By discouraging quality-sensitive contributors, platforms risk degrading the very content that sustains long-term engagement. Managers should therefore calibrate advertising levels to strike a balance between short-term revenue and preserving the high-quality user base that drives competitive advantage. Second, the model shows that first-mover advantage can be decisive: once a platform secures a dominant position and commits to a sustainable advertising strategy, rivals may find entry prohibitively difficult. This suggests that early investment in building a loyal, quality-sensitive user base can yield durable dominance. Finally, platform operators should recognize that user heterogeneity creates opportunities for differentiation; smaller entrants may succeed by cultivating communities that prize quality over scale, rather than competing head-to-head on network size.

### *5.5. Policy Implications*

From a policy perspective, the model raises concerns about market concentration. When advertising intensity is used strategically, it can reinforce tipping and lead to monopoly-like outcomes. This creates risks of reduced innovation, diminished user choice, and potential exploitation once rivals are excluded. Regulators should therefore pay attention not only to traditional measures of market power, such as pricing, but also to the ways in which platforms shape user composition and content quality. Policies that encourage interoperability, data portability, or limits on excessive advertising may help maintain a healthier competitive environment.

### *5.6. Potential Limitations*

Our model assumes symmetric cost structures and purely advertising-based revenue. The solution is relatively robust to asymmetric cost structures: the profit of platform 2 would still be driven to zero.

Meanwhile, alternative monetization schemes may also alter equilibrium outcomes. Under a freemium or subscription model, paying users are insulated from advertising, which weakens the negative link between advertising intensity and perceived quality. In such settings, platforms could extract revenue without driving away quality-sensitive contributors, potentially softening the collapse into monopoly that emerges in our baseline. Similarly, hybrid models combining ads and subscriptions would generate richer strategic trade-offs, where platforms balance ad loads against the retention of premium users.

These considerations suggest that while our baseline model captures important dynamics of ad-funded platforms, incorporating cost heterogeneity and alternative revenue models would yield a broader understanding of competitive strategies in practice.

## 6. Conclusion

This paper developed a theoretical model of digital platform competition in which user-generated content quality is an endogenous outcome driven by user heterogeneity and strategic advertising decisions. We extended prior work by modeling content quality not as a fixed attribute but as a function of the platform's user base composition, itself determined by equilibrium platform choice. Our analysis shows that while platforms can compete by adjusting advertising intensity, doing so also alters the perceived content quality, influencing user distribution and platform profitability.

The equilibrium analysis reveals a trade-off between attracting a broad user base and maintaining high content quality. When users vary significantly in their relative preferences for quality and network effects, platforms face nontrivial incentives that can lead to multiple equilibria. However, as we have shown, simultaneous competition can produce a destructive collapse when each platform tries to drive the other's profit to zero, whereas sequential play allows a leader to preempt the follower and effectively hoard the market.

Overall, this paper sheds light on how quality-sensitive users and platform decisions interact to shape the competitive landscape. Future extensions could consider dynamic user engagement, platform investment in moderation or curation, or empirical calibration to specific platform data.

Future research might explore some natural extensions of the model, for example, incorporating dynamic user behavior, where users update their beliefs or switch platforms over time based on evolving perceptions of quality and network size. Another direction is to introduce platform-side investments in quality moderation or algorithmic promotion, which could affect not only perceived quality but also the endogenous composition of the user base. For example, platforms might actively discourage low-quality contributions or offer tools to amplify high-quality ones, thereby influencing equilibrium outcomes.

Finally, while our model focuses on symmetric cost structures and purely advertising-driven revenue, incorporating heterogeneous cost functions or alternative monetization models (such as subscriptions or freemium pricing) could reveal new strategic trade-offs. These extensions would deepen our understanding of how digital platforms navigate the balance between monetization, quality, and user engagement in competitive environments.